\documentstyle[12pt]{article}
\newcommand{\bea}{\begin{eqnarray}}
\newcommand{\eea}{\end{eqnarray}}

\newcommand{\ee}{\end{equation}}
\newcommand{\be}{\begin{equation}}

\begin{document}
\title{
Weak-QES extensions of the Calogero model}
\author{{\large Y. Brihaye}$^{\diamond}$
and {\large P. Kosinski}$^{\dagger}$
\\
$^{\diamond}${\small Dep. of Mathematical Physics, University of Mons-Hainaut,
Mons, Belgium}\\
$^{\dagger}${\small 
Dep. of Theoretical Physics, University of Lodz, Lodz, Poland}}
\date{}
\maketitle
\begin{abstract}
We construct families of Hamiltonians  extending the Calogero model and
such that a finite number of eigenvectors can be computed algebraically.
\end{abstract}
\medskip
\medskip
\section{Introduction}
The notion of quasi exactly solvable equations 
has many different meanings. 
Initially \cite{tur,ush} it was invented to
qualify operators (typically quantum Hamiltonian ones) which,
after suitable change of variables and(or) "gauge rotation",
are equivalent to an element of the envelopping algebra of some Lie
algebra, represented by differential operators.
In the following we refer to the above property as to algebraic
 QES property.
In opposition some operators can possess a series of values of the 
coupling constants for which an explicit eigenvector is available
without beeing related to any representation of a Lie algebra.
We will qualify this property as "weak-QES". 
In the recent few years the Calogero model
\cite{cal} (and several of its
extensions \cite{per}) 
have received a considerable new impetus of interest.
One of the new results was the construction \cite{rutu}
 of a set of variables (the so called $\tau$-variables)
in which the N-body Calogero model can be written as an element
of the enveloping algebra of $sl(N)$. 
The complete integrability of the model was then directly related
to the finite representations of a Lie algebra, following very closely
one of the ideas underlying the notion of quasi exact solvability 
\cite{tur}. 
Some algebraic QES generalisations of the Calogero models were
proposed soon after the basic result of \cite{rutu}.
Unfortunately these extentions are all based of an $sl(2)$
algebra.  Basically, only one of the coordinates,
for instance $\tau_2$, is involved
into the additional piece of the potential.
The purpose of this note is to exhibit
series of weak-QES hamiltonians, generalizing the Calogero
hamiltonian, some of them depending generically of all the $\tau$'s.

\section{The Hamiltonian}

We are interested in hamiltonians of the form
\be
\label{cal}
H = H_c + V
\ee
\be
\label{potcal}
H_c = {1\over 2} (-\Delta + \omega^2 r^2) + \sum^N_{j<k=1}
{\nu(\nu-1)\over{(x_j-x_k)^2}}
\ee
with
\be
\label{lapla}
\Delta = \sum^N_{i=1} {\partial^2\over{\partial x^2_i}} \quad ,
\quad r^2 = \sum^N_{i=1} x^2_i  
\ee
and such that the eigenvalue equation
\be
 H\psi = E\psi
\ee
admits eigenvectors of the form
\be
\psi(x) = (\beta(x))^{\nu} e^{-{\omega\over 2}r^2} e^{P(x)}p(x)
\ \ , \ \ \beta(x) = \Pi_{j<k}(x_j-x_k) \ \ ,
\ee
The function $\beta(x)$ is the Vandermonde determinant of the matrix
$M_{ij} = (x_i)^j$. 
Moreover we restrict ourselves to polynomial forms
of $V(x), P(x), p(x)$ in $x\equiv (x_1,x_2, \dots ,x_N)$.
The operator acting on $p(x)$ 
will be denoted  $\tilde H$ :
\be
\label{calred}
\tilde Hp = Ep
\ee

In the following we  use the variables 
$\sigma_1, \tau_2,\tau_3,\cdots,\tau_N$ 
introduced in \cite{rutu}.
In these variables,  the laplacian 
(\ref{lapla}) reads \cite{rutu}
\be
\Delta(\tau) = N{\partial^2\over{\partial \sigma^2_1}} + \sum^N_{j,k=2}
A_{jk} {\partial^2\over{\partial \tau_j\partial \tau_k}} + \sum^N_{j=2}
B_j {\partial \over{\partial \tau_j}}
\ee
\begin{eqnarray}
A_{jk} &=& {(N-j+1)(k-1)\over N} \tau_{j-1} \tau_{k-1}\nonumber\\
&+& \sum_{l \geq \max (1,k-j)} (k-j-2l) \tau_{j+l-1} \tau_{k-l-1}\\
B_j &=& -{(N-i+2)(N-i+i)\over N} \tau_{i -2}
\end{eqnarray}
For the manipulation of a 
generic (translation-invariant) change of basis, the relevant 
formula reads 
\be
\sum_k {\partial w\over{\partial x_k}} {\partial\over{\partial x_k}} =
\sum^N_{j,k=2} A_{jk} {\partial w\over{\partial \tau_j}} {\partial\over
{\partial \tau_k}}
\ee
The following identities are also useful~:
\be
\beta^{-\nu} (-{1\over 2} \Delta(x) + \sum_{j<k} {\nu(\nu-1)\over{(x_j-x_k)^2}})
\beta^{\nu}
= -{1\over 2} \Delta(\tau) + {\nu\over 2} \sum^N_{j=2} (N-j+2) (N-j+1)
\tau_{j-2} {\partial\over{\partial \tau_j}}
\ee
\be
e^{{\omega\over 2}r^2} (-{1\over 2} \Delta(x) + {\omega^2\over 2} r^2)
e^{-{\omega\over 2}r^2} = -{1\over 2} \Delta (\tau) + \omega
\sum^N_{j-2} j \tau_j {\partial\over{\partial \tau_j}} + {N\over 2}\omega
\ \ \ ;
\ee
they allow to handle respectively
the repulsive and the harmonic parts of the
potential in (\ref{cal}).

\par We have attempted to construct the
polynomial potentials $V(\tau_2,\tau_3,\cdots,\tau_N)$
such that the operator (\ref{cal}) is equivalent (after a change 
of basis and
with the variables $\tau$) to an operator preserving 
one vector space of the form
\be
\label{space}
{\cal P}(N,n) = {\rm span}
 \lbrace \tau_2^{n_2} \tau^{n_3}_3\cdots \tau^{n_N}_N
 \ | \ n_2+n_3
+\cdots + n_N \leq n\rbrace
\ee
Our calculations indicate that, for $N=2,3,4,5$, the
Calogero hamiltonian (corresponding to $V=0$ in (\ref{cal}))
is the only one to possess such a 
property. After this negative result, we investigate
alternative possibilities of algebraic solutions 
by imposing weaker requirements.
Four types of situation have been considered which
presented in the next section.

\section{Weak-QES Hamiltonians}

\subsection{Type 1}

\par We set $N=3$; as previously stated $V,P,p$ 
are then polynomials in $\tau_2, \tau_3$.
Use of the identities of Sec. 2 leads to the operator 
$\tilde H$ acting on  $p(\tau_2, \tau_3)$ ~:
\begin{eqnarray}
\label{calre4}
\tilde H &=& \tau_2 {\partial^2\over{\partial \tau^2_2}} + 3\tau_3
{\partial^2\over{\partial \tau_2 \partial \tau_3}} - {1\over 3} \tau^2_2
{\partial\over{\partial \tau^2_3}} + {\partial\over{\partial \tau_2}}\nonumber\\
&+& ({\partial P\over{\partial \tau_2}}) (2\tau_2 {\partial\over{\partial
\tau_2}} + 3\tau_3 {\partial\over{\partial \tau_3}})\nonumber\\
&+& ({\partial P\over{\partial \tau_3}}) (3\tau_3{\partial\over{\partial \tau_2}}
- {2\over 3} \tau^2_2 {\partial\over{\partial \tau_3}})\nonumber\\
&+& \omega (2\tau_2 {\partial\over{\partial \tau_2}} + 3\tau_3 {\partial
\over{\partial \tau_3}})\nonumber\\
&+& 3\nu {\partial \over{\partial \tau_2}} + V_{eff}
\end{eqnarray}
where we define
\be
V_{eff} \equiv V-{1\over 2} e^{-P} \Delta e^P
\ee
We have constructed the solutions of this equation  for
particular values of the degrees of $P$  and of $p$. 

\noindent {\bf{P \ is \ of \ degree \ four}}

We assume $P(x)$ to be at most quartic in $x$, i.e.
\be
P = {c\over 2} \tau^2_2 + b \tau_2 + d\tau_3
\ee
Choosing for $p$ a polynomial of global degree $n$ in $\tau_2, \tau_3$,
a careful power counting in (\ref{calred}), (\ref{calre4}),
reveals that polynomial solutions can exist 
only if  $V_{eff}$ is the form
\be
V_{eff} = v_1 \tau_2 + v_0
\ee
($v_0$ accounts for the eigenvalue of $\tilde H$, i.e. $v_0 \equiv -E$).
Moreover  there are generically
${(n+2)(n+3)\over{2}} - 1$ algebraic equations to be solved and 
${(n+1)(n+2)\over 2}+1$ parameters 
(including the parameters of $p$ and the two defining $V_{eff}$).
We solved explicitely these equations for the first few 
values of $n$.
\par Let $n=1$, then
\be
p = \tau_2 + c_1 \tau_3 + c_0
\ee
Two types of solutions are found after some algebra.

\vspace{0.5cm}

{\large {Solution (a)}}
\begin{eqnarray}
c_1 &=& d = 0\qquad , \qquad v_1 = -2c\nonumber\\
c_0 &=& {2b+2\omega+v_0\over {2c}}\nonumber\\
v_0 &=& -b-w \pm \sqrt{\omega^2-2c-6c\nu + 2b \omega + b^2}
\end{eqnarray}
This solution does not depend on $\tau_3$
and is therefore not generic. It depends only on $\tau_2$
and it has two possible of eigenvalues of the energy ($E\equiv -v_0$). 
It is a particular case of the QES extension of the Calogero
model constructed in \cite{min}. The potential depends on four parameters
$\omega,\nu,b,c$.  Such solutions can be constructed for any values
of $n$. The corresponding Hamiltonian is an element of the 
enveloping algebra of sl(2) realized on the space of polynomials of degree
at most $n$ in $\tau_2$.

\vspace{0.5cm}

{\large{Solution (b)}}
\begin{eqnarray}
c_1 &=& {-3c\over{2d}}\qquad , \qquad c_0 = {2d^2-\omega c-bc\over{3c^2}}\nonumber\\
v_1 &=& -3c \qquad , \qquad v_0 = {1\over c} (2d^2-3\omega c-3bc)
\end{eqnarray}
which has to be supplemented by one condition on $\omega, \nu, b,c,d$:
\be
\label{rel}
3 b^2 c^2+6 b c^2 \omega - 8 b c d^2
+ 9 c^3 \nu + 3 c^3 + 3 c^2 \omega^2
- 8 c d^2\omega + 4 d^4
= 0
\ee
Unlike the "solution (a)", 
this solution non trivially depends on $\tau_2$ and $\tau_3$. 
The potential is parametrized by 
$\omega,\nu, b,c,d$ constrained by (\ref{rel}). 
One can in principle solve (\ref{rel}) with respect to one of the
constants and obtain a finite number of potentials each admitting one
algebraic eingenvector. This result is similar in spirit
to a QES-type II equation \cite{tur}. 

Repeating the above calculation for $n=2$, i.e.
\be
p = \tau^2_2 + c_1 \tau_2\tau_3+ c_3\tau^2_3 +c_2 \tau_2 + c_4 \tau_3 + c_5
\ee
As a counterpart of solution (b), we find the following expressions
\be
v_1 =  - 6 c \ \ , \ \ 
v_0 = 2{  2 d^2 - 3 b\ c - 3 c \ \omega  \over c}
\ee

\be
c_1 = -3 {c \over d} \ \ , \ \ 
c_2 = {32 d^4 - 16\ b\ c\ d^2 - 9 c^3 - 16 c\ d^2 \omega  
\over 24 c^2\ d^2}
\ee

\be
c_3 = {9 c^2 \over 4 d^2} \ \ , \ \ 
c_4 = {b \ c + c \ \omega - 2 d^2 \over c \ d}
\ee

\bea
c_5 &=& (16 b^2\ c^2\ d^2 + 9 b\ c^4 + 32 b\ c^2\ d^2\ \omega 
- 48 b\ c\ d^4 + 9 c^4 \omega  \nonumber \\ 
&+& 36 c^3 d^2 \ \nu + 15 c^3\ d^2 + 16 c^2\  d^2 \omega^2 
- 48 c\ d^4 \omega + 32 d^6)/(36 c^4 d^2)
\eea
which fix the parameters of $p$ and $V_{eff}$ in 
terms of the coupling constant $\omega , \nu, b,c,d$. 
Two supplementary relations, 
analog to (\ref{rel}), have to be imposed
of these coupling constants
\be
0 =  24 b^2 c^2 + 48 b c^2 \omega - 64 b c d^2 + 72 c^3 \nu
 + 105 c^3 + 24 c^2 \omega^2 - 64 c d^2 \omega + 32 d^4 
\ee

\bea
0 &=& - 192 b^3 c^3 d^2 - 108 b^2 c^5 
      - 576 b^2 c^3 d^2 \omega + 704 b^2 c^2 d^4 \nonumber \\
   &-& 216 b c^5 \omega - 576 b c^4 d^2 \nu 
- 156 b c^4 d^2 - 576 b c^3 d^2 \omega^2       \nonumber  \\
&+& 1408 b c^2 d^4 \omega - 768 b c d^6 
- 81 c^6 \ \nu - 27 c^6 - 108 c^5 \ \omega^2 
- 576 c^4\ d^2\ \omega \ \nu  
\nonumber  \\
&-& 156 c^4 d^2 \omega + 576 c^3\ d^4 \nu + 216 c^3\ d^4
 - 192 c^3\ d^2 \omega^3 + 704 c^2\ d^4 \omega^2 \nonumber \\
&-& 768 c\ d^6 \omega + 256 d^8 
\eea
So that we end up with a three-parameters family of 
weak-QES potentials. 

\noindent {\bf{P \ is \ of \ degree \ six}}

We also considered the case of a
polynomial $P$ of degree at most six in $x$, i.e.
\be
P = {c\over 2} \tau^2_2 + b \tau_3 + d \tau_2 + {\tilde c\over 3} \tau^3_2
+ {\tilde b\over 2} \tau^2_3 + \tilde d \tau_2\tau_3
\ee
and found that  $V_{eff}$ has to be of the form
\be
    V_{eff} = v_0 + v_1 \tau_2 + v_2 \tau_3 + v_3 \tau_2^2
\ee
i.e. at most quartic in the variables $x_i$.
Solving the equations  (\ref{calred}) for $n=3$, we checked that 
a four-parameter family of weak-QES potentials exists.
The form of the constraints on the parameters rapidly
becomes cumbersome when the degree of $P$ increases.

\vspace{0.5cm}

\subsection{Type 2}

\par 
When we demand the functions $V, P, p$
in (1)-(5) to depend on the variable 
$\tau_2$ only,  the
operator $\tilde H$ takes the form
\begin{eqnarray}
\tilde H &=& \tau_2 {\partial^2\over{\partial \tau^2_2}} + (\tau_2
P' + 2\omega \tau_2 + {N-1\over 2} (1+N\nu)){\partial\over{\partial \tau_2}}
\nonumber\\
&+& (2\omega \tau_2 + {N-1\over 2} (1+N\nu))P' + \tau_2 (P''+ (P')^2)+V(\tau_2)
\end{eqnarray}
where $P' = {dP\over{d\tau_2}}$, etc.
\par The case $P = -{a\over 2} \tau^2_2 - b \tau_2$ corresponds to the QES
extension of \cite{min};
the potential $V$ is then of third degree in $\tau_2$.  However
weak-QES potentials of higher degree can be constructed.
 Let indeed $V$ be of degree $2 \delta+1$,
then a power counting reveals that $P$ has to be of degree $\delta+1$
and that the algebraic equation (\ref{calred})
can be fulfilled with a polynomial $p(\tau_2)$ 
provided a number of  $\delta$ conditions among the $2 \delta+1$
coupling constants entering in the potential are satisfied.
This results into families of weak-QES  potentials depending of $\delta+1$ parameters.

\subsection{Type 3}

\par We have reconsidered the QES
Hamiltonian of \cite{sh} and tried to generalize it
by following the approach presented in Sec.2. 
The hamiltonian   has the form  \cite{sh} 
\be
H = -\Delta + V + g \sum^N_{i,j=1,i\not=j} 
\bigl( {1\over{(x_i-x_j)^2}} -
{1\over{(x_i+x_j)^2}} \bigr) \ \ .
\ee
We look for the most general change 
of function $e^P$ and potential $V$
such that
\be
\tilde H = e^P H e^{-P}|_{\xi_i}\qquad , <qquad \xi_i \equiv x^2_i
\ee
is an element of the enveloping algebra of the Lie
algebra $sl(N)$ in the representation given by 
Eq.(16) of \cite{rutu}. After an algebra, we find a four parameters family
of potentials with
\be
e^{-P} = (\Pi^N_{i,j=1,i<j} (\xi_i-\xi_j))^{\alpha} (\Pi^N_{j=1} \xi_j)^{\beta}
\exp \lbrace -{a\over 4} \sum^N_{j=1} \xi^2_j - {b\over 2} \sum^N_{j=1}
\xi_j\rbrace
\ee
and
\be
V = \sum^N_{i=1}\lbrace {g'\over{4\xi_i}} + (b^2-a(4n+4\alpha(N-1) + 4\beta
+ 3)) \xi_i  + 2ab \xi^2_i + a^2\xi^3_i\rbrace
\ee
with $g  = \alpha(\alpha-1), g' = 2\beta (2\beta-1)
$. This is slightly more general than
the potential given in  \cite{sh} since the coupling
constants $g$ and $g'$ are here independent.

\vspace{0.5cm}
\subsection{Type 4}

\par Finally we present a result which  directly generalizes to
N dimensions the famous
one-dimensional  sextic QES potential \cite{tur,ush}.
In this purpose we consider
\be
H = -{1\over 2}\Delta + V_6 (x)
\ee
and assume that the even function 
$V_6(x)$  contain powers of degree at most six in $x_j$.
The most general change of function $e^P$ such that
the operator
\be
\tilde H = e^P H e^{-P}|_{t_i}\ \  \ , \ \ \  t_i = x_i^2
\ee
is an element of the enveloping algebra of $sl(N)$ 
(in the representation
given by Eq.(16) 
of \cite{rutu}) is given by the $2N+1$-parameters function
\be
P = \alpha T^2 + 
\sum^N_{a=1} p_a t_a + 
\sum^N_{a=1} q_a \log t_a\quad , \quad
T = \sum^N_{i=1} t_i
\ee
where $\alpha, p_a, q_a$ are constants. Correspondingly the most general
QES potential reads         
\be
V_6 = 4\sum_i t_i ({\partial P\over{\partial t_i}})^2 + 4\sum_i t_i
{\partial^2P\over{\partial t_i^2}} + 2\sum_i {\partial P\over{\partial t_i}}
- 16\alpha n T
\ee
For $N=2$, the 2-body 
polynomial potential of \cite{ush, shtu} is then recovered as
the special case $q_1 = q_2 = 0$.

Summarizing, we have analyzed the rational Calogero
model from the point of view of the notion of quasi exact 
solvability. We found, at least for the small number of particles,
that it is unique in the following sense~:
it is exactly solvable but has no exactly or quasi-exactly solvable
translational-invariant extension except those following from an
sl(2) structure. By relaxing the notion of quasi-exact solvability
(weak QES) we were able to find a number of extensions of the Calogero
model. They are characterized by the existence of analytical expressions
for some levels without hidden symmetry behind.

\bigskip
{\bf Acknowledgments.} This work was
supported by the Belgian F.N.R.S. and
carried out under 
the grant n$^0$ 2 P03B 076 10 of the Polish Government.   


\end{document}